\documentclass[twocolumn,aps,prc,superscriptaddress,nofootinbib,showpacs,floatfix,longbibliography]{revtex4}
\usepackage{url}
\usepackage{cancel}
\usepackage[colorlinks,linkcolor=blue,citecolor=blue,filecolor=black,urlcolor=blue]{hyperref}
\usepackage{epsfig,graphics}
\usepackage{graphicx}
\usepackage{dcolumn}
\usepackage{bm}
\usepackage{amssymb}
\usepackage{amsmath}
\usepackage{multirow}
\usepackage{float}
\usepackage{harpoon}
\usepackage{MnSymbol}
\usepackage{appendix}
\usepackage{color}
\usepackage{hyperref}
\usepackage{cleveref}

\newcommand{\nch} {N_{\mathrm{ch}}}
\newcommand{\sqrtsnn}{\mbox{$\sqrt{s_{\mathrm{NN}}}$}}
\newcommand{\pT} {p_{\mathrm{T}}}
\newcommand{\lr}[1]{\left\langle #1\right\rangle}

\newcommand{\ruru}{$^{96}$Ru+$^{96}$Ru}
\newcommand{\zrzr}{$^{96}$Zr+$^{96}$Zr}

\begin{document}
\title{Ratios of collective flow observables in high-energy isobar collisions are insensitive to final state interactions}
\newcommand{\bnl}{Physics Department, Brookhaven National Laboratory, Upton, NY 11976, USA}
\newcommand{\sbu}{Department of Chemistry, Stony Brook University, Stony Brook, NY 11794, USA}
   \author{Chunjian Zhang}\email{chun-jian.zhang@stonybrook.edu}\affiliation{\sbu}
   \author{Somadutta Bhatta}\affiliation{\sbu}
   \author{Jiangyong Jia}\email{jiangyong.jia@stonybrook.edu}\affiliation{\sbu}\affiliation{\bnl}
  \date{\today}
 
\begin{abstract}
The ratios of bulk observables, such as harmonic flow $v_2$ and $v_3$, between high-energy $^{96}$Ru+$^{96}$Ru and $^{96}$Zr+$^{96}$Zr collisions were recently argued to be a clean probe of the nuclear structure differences between $^{96}$Ru and $^{96}$Zr. Using a transport model simulation of isobar collisions, we quantify this claim from the dependence of the ratios $v_{2,\mathrm{Ru}}/v_{2,\mathrm{Zr}}$ and $v_{3,\mathrm{Ru}}/v_{3,\mathrm{Zr}}$ on various final state effects, such as the shear viscosity, hadronization and hadronic cascade. Although the $v_2$ and $v_3$ change by more than 50\% when varying the final state effects, the ratios are unchanged. In addition, these ratios are independent of the transverse momentum $p_{\mathrm{T}}$ and hadron species, despite of up to a factor of two change in $v_n$. The ratio of mean transverse momentum $\left\langle p_{\mathrm{T}}\right\rangle$ is found to be controlled by the nuclear skin and nuclear radius, but is only slightly impacted by the final state effects. Therefore, these isobar ratios serve as a clean probe of the initial condition of the quark-gluon plasma, which in turn is controlled by the collective structure of the colliding nuclei.
\end{abstract}
\maketitle

\textit{\bf Introduction.} The success of the hydrodynamic framework of heavy-ion collisions enabled quantitative extractions of the transport properties of the quark-gluon plasma (QGP) as for instance done in the state-of-the-art multi-system Bayesian analysis approaches~\cite{Bernhard:2016tnd,Everett:2020xug,Nijs:2020ors}. Such extractions rely on a correct description of the initial condition of the QGP prior to the hydrodynamic expansion. The initial condition, e.g. its energy density distribution, is generated from the two colliding nuclei with simple scaling ansazes in Glauber model or Glauber-inspired models, which often assume spherical nuclei described by a two-parameter Fermi function~\cite{Miller:2007ri,Moreland:2014oya}. Recent experimental data~\cite{Adamczyk:2015obl,Acharya:2018ihu,Sirunyan:2019wqp,Aad:2019xmh,ALICE:2021gxt,ATLAS:2022dov} and dedicated theoretical studies~\cite{Heinz:2004ir,Shou:2014eya, Giacalone:2019pca,Li:2019kkh, Giacalone:2021uhj,Jia:2021tzt, Giacalone:2021udy, Jia:2021qyu,Jia:2021wbq,Zhang:2021kxj,Xu:2021uar,Bally:2021qys,Jia:2021oyt,Nijs:2021kvn,Zhao:2022uhl}, however, have indicated the importance of nuclear deformations and nuclear radial distributions of proton and neutrons in the nucleus, on the measured multiplicity distribution and collective flow.  For a precision study of the properties and initial condition of QGP, the impact of and relations to these collective nuclear structure effects should be quantified, especially since all species used in high-energy nuclear collisions are expected to present some deformations and some uncertainties in their radial profiles. 

The best example highlighting the importance of collective nuclear structure, is offered by the measurements in \ruru\ and \zrzr\ collisions at $\sqrtsnn=200$ GeV by the STAR Collaboration at RHIC~\cite{STAR:2021mii}. Ratios of many observables between $^{96}$Ru+$^{96}$Ru and $^{96}$Zr+$^{96}$Zr are found to deviate from unity in an observable- and centrality-dependent manner, including anisotropic flow $v_2$, $v_3$ and charged particle multiplicity $\nch$. These deviations must first originate from structure differences between $^{96}$Ru and $^{96}$Zr, which impact the initial condition of QGP and its final state observables. Indeed, model comparison to the published $v_n$ and $\nch$ data reveals a larger quadrupole deformation $\beta_2$ in $^{96}$Ru, a larger octupole deformation $\beta_3$ in $^{96}$Zr, and a larger diffusiveness value consistent with a larger neutron skin in $^{96}$Zr~\cite{Jia:2021oyt,Nijs:2021kvn}. Predictions for many other observables and their sensitivities to deformation and neutron skin have been made, such as mean transverse momentum $\lr{\pT}$~\cite{Xu:2021uar} and its fluctuations~\cite{Jia:2021qyu}, spectator neutron production~\cite{Liu:2022kvz}, mixed-flow harmonics~\cite{Zhao:2022uhl,Jia:2022qrq}, and $v_n$--$\pT$ correlations~\cite{Giacalone:2019pca,Bally:2021qys,Jia:2021wbq}. Preliminary measurements on a subset of these observables recently became available, which largely confirm these model predictions~\cite{chunjianhaojie}. All these progresses highlight the enormous potential of using isobar collisions to improve our understanding of the initial condition of heavy ion collisions. 

Before drawing a strong conclusion, an important question must be addressed first: ``{\it To what extent are the isobar ratios insensitive to the final state effects, including medium properties, particlization and hadronic transport?}''. To partially address this question, we performed a detailed investigation of the impact of the shear viscosity $\eta/s$ and hadronic transport on the ratios of several observables in isobar collisions within a transport model. Our findings support the idea that the isobar ratios mainly reflect the initial condition. 

\textit{\bf Setup.} We simulate isobar collisions using the popular AMPT transport model~\cite{Lin:2004en}, which starts with fluctuating initial conditions from a Monte Carlo Glauber model. The system evolution is modeled with strings that first melt into partons, followed by elastic partonic scatterings, parton coalescence, and hadronic scatterings. The hydrodynamic collectivity is generated mainly through elastic scatterings of partons. The specific shear viscosity $\eta/s$ is controlled by the partonic cross-section $\sigma=9/2 \pi \alpha_{s}^{2}/\mu^{2}$, where $\alpha_s$ is the QCD coupling constant and $\mu$ is the screening mass. For a system of two massless quark flavors at initial temperature $T$, the $\eta/s$ can be estimated using the following pocket formula~\cite{Xu:2011fi},
\begin{eqnarray}\label{eq:1}
\frac{\eta}{s}\approx\frac{3 \pi}{40 \alpha_{s}^{2}} \frac{1}{\left(9+\frac{\mu^{2}}{T^{2}}\right) \ln \left(\frac{18+\mu^{2} / T^{2}}{\mu^{2} / T^{2}}\right)-18}
\end{eqnarray}

Following our previous work~\cite{Jia:2021oyt}, we use the AMPT model v2.26t5 with string-melting mode and hadronic cascade time $\tau_{hc}=$ 30 fm/$c$. Four different combinations of $\alpha_s$ and $\mu$ are chosen, as listed in Table~\ref{tab:1}, to give different partonic cross-section of 1.5, 3.0, 6.0 and 10.0 $m$b, respectively. Among these, the default case of 3.0~$m$b has been shown to reproduce the Au+Au $v_2$ data at RHIC~\cite{Ma:2014xfa}. Assuming initial temperature of $T=0.38$~GeV, it would correspond to $\eta/s$ value of 0.23 at early time. Correspondingly, the case of 1.5 $m$b gives a larger $\eta/s$ value, while the cases of 6.0 and 10.0 $m$b give smaller $\eta/s$ value~\cite{Xu:2011fi}. The impact of the hadronic scattering is studied by reducing the hadronic cascade time from 30 fm/$c$ to 15 fm/$c$. Note that the QGP is expected to spent more time at a lower temperature where the $\eta/s$ value is smaller (however the perturbative formula Eq.~\ref{eq:1} may not be applicable), and therefore the $\eta/s$ values in Table~\ref{tab:1} should be treated as a crude estimate. Note as well that varying partonic cross-section might also change the specific bulk viscosity, which do not impact $v_n$ but may influence $\lr{\pT}$ via radial flow. However, the exact values of shear and bulk viscosities are irrelevant for our study, what is important is that these choices significantly change the predicted $v_n$, and therefore allow us to check the stability of the isobar ratios. 

\begin{table}[h!]
\caption{\label{tab:1} The parameter settings in the AMPT model, where the $\eta/s$ value is estimated from Eq.~\eqref{eq:1} assuming an initial temperature $T=0.38$~GeV. The last two rows list the Woods-Saxon parameter values used for isobar collisions via Eq.~\eqref{eq:2}.}
\begin{ruledtabular}
\begin{tabular}{lcccc}
$\alpha_s$ & $\mu$ (fm$^{-1}$)& $\sigma$ ($m$b) &$\eta/s$ (Eq.~\eqref{eq:1}) & $\tau_{hc}$ (fm/$c$)\\
 0.33 & 3.226 & 1.5 & 0.387& 30 \\
 0.33 & 2.265 & 3.0 & 0.232& 30 \\
 0.33 & 2.265 & 3.0 & 0.232& 15 \\
 0.33 & 1.602 & 6.0 & 0.156& 30 \\
 0.48 & 1.800 & 10  & 0.087& 30 \\\hline\hline                          
Species  &\; $R_0$ (fm)\; & \;$a_{0}$ (fm)\;  & $\beta_{2}$ & $\beta_{3}$ \\
$^{96}$Ru & 5.09  & 0.46   & 0.162 & 0 \\
$^{96}$Zr & 5.02  & 0.52   & 0.06 & 0.20 \\
\end{tabular}
\end{ruledtabular}
\end{table}

The spatial distribution of nucleons in the $^{96}$Ru and $^{96}$Zr nuclei is described by a deformed Woods-Saxon (WS) function,
\begin{align}\label{eq:2}
\rho(r,\theta,\phi)&\propto\frac{1}{1+e^{[r-R_0\left(1+\beta_2 Y_2^0(\theta,\phi) +\beta_3 Y_3^0(\theta,\phi)\right)]/a_0}},
\end{align}
with four parameters: quadrupole deformation $\beta_2$, octupole deformation $\beta_3$, surface diffuseness $a_0$ and half-density nuclear radius $R_0$. The parameter values for $^{96}$Ru and $^{96}$Zr are chosen to be identical as our previous studies ~\cite{Zhang:2021kxj,Jia:2021oyt} and they are listed in Table~\ref{tab:1}. In addition, we also simulate three intermediate cases whose parameter values are chosen such that we can separate the influence of each parameter. 

The flow harmonics $v_n$ for $n=2$ and 3 are calculated from the generated AMPT events using the standard two-particle correlation method,
\begin{eqnarray}\label{eq:3}
\left\langle v_{n}^2\right\rangle=\left\langle\left\langle e^{in(\phi_i-\phi_j)}\right\rangle\right\rangle\;,
\end{eqnarray}
where $\left\langle\left\langle \right\rangle\right\rangle$ represents the average over unique pairs in each event, and then over all events in an event class. To enhance the statistical precision, all hadrons within the pseudorapidity range $|\eta|<2$ and transverse momentum range $0.2<\pT<2$ GeV/$c$ are used. These particles are also used to calculate the average transverse momentum $\lr{\pT}$. Following the STAR data analysis~\cite{STAR:2021mii}, $\nch$ in each event is defined as the number of charged particles within $|\eta|<0.5$ and $\pT>0.1$ GeV/$c$, from which we obtain the probability distribution $p(\nch)$. The impact of short-range non-flow correlations are studied by reporting the $\left\langle v_{n}^2\right\rangle$ calculation using two-subevent method, i.e. by correlating particles in $\eta>0$ with those in $\eta<0$. Almost all results look the same but with larger statistical uncertainties. Therefore the default results are obtained without $\Delta\eta$-gap, only in the case of PID-dependence where some differences are observed, the two-subevent method is chosen as the default method. 

For an observable $\mathcal{O}$, the isobar ratio is calculated at the matching $\nch$ between \ruru\ and \zrzr\ collisions,
\begin{align}\label{eq:4}
R_{\mathcal{O}}(\nch)=\frac{\mathcal{O}_{\mathrm{Ru}}(\nch)}{\mathcal{O}_{\mathrm{Zr}}(\nch)}\;.
\end{align}
Note that this ratio is slightly different from those calculated as the matching centrality due to the fact that the $p(\nch)$ is also slightly different between the two collision systems~\cite{Jia:2022iji}.  Our previous study shows that isobar ratios are controlled by the differences in the nuclear structure parameters~\cite{Zhang:2021kxj},
\begin{align}\label{eq:5}
R_{\mathcal{O}}\approx 1+ c_1 \Delta\beta_2^2 +c_2 \Delta\beta_3^2 + c_3\Delta R_0 +c_4\Delta a_0\;,
\end{align}
with $\Delta \beta_2^2 = \beta_{\mathrm{2,Ru}}^2-\beta_{\mathrm{2,Zr}}^2$, $\Delta \beta_3^2 = \beta_{\mathrm{3,Ru}}^2-\beta_{\mathrm{3,Zr}}^2$, $\Delta a_0 =a_{\mathrm{0,Ru}}-a_{\mathrm{0,Zr}}$ and $\Delta R_0 =R_{\mathrm{0,Ru}}-R_{\mathrm{0,Zr}}$. The coefficients $c_1$--$c_4$ encode the information about how the energy from the colliding nuclei is deposited to create initial condition of QGP. We shall show that the $R_{\mathcal{O}}$ are insensitive to the final state effects, and therefore the $c_n$ are robust probe of the initial condition of heavy ion collisions.
\begin{figure*}[!ht]
\centering
\includegraphics[width=0.7\linewidth]{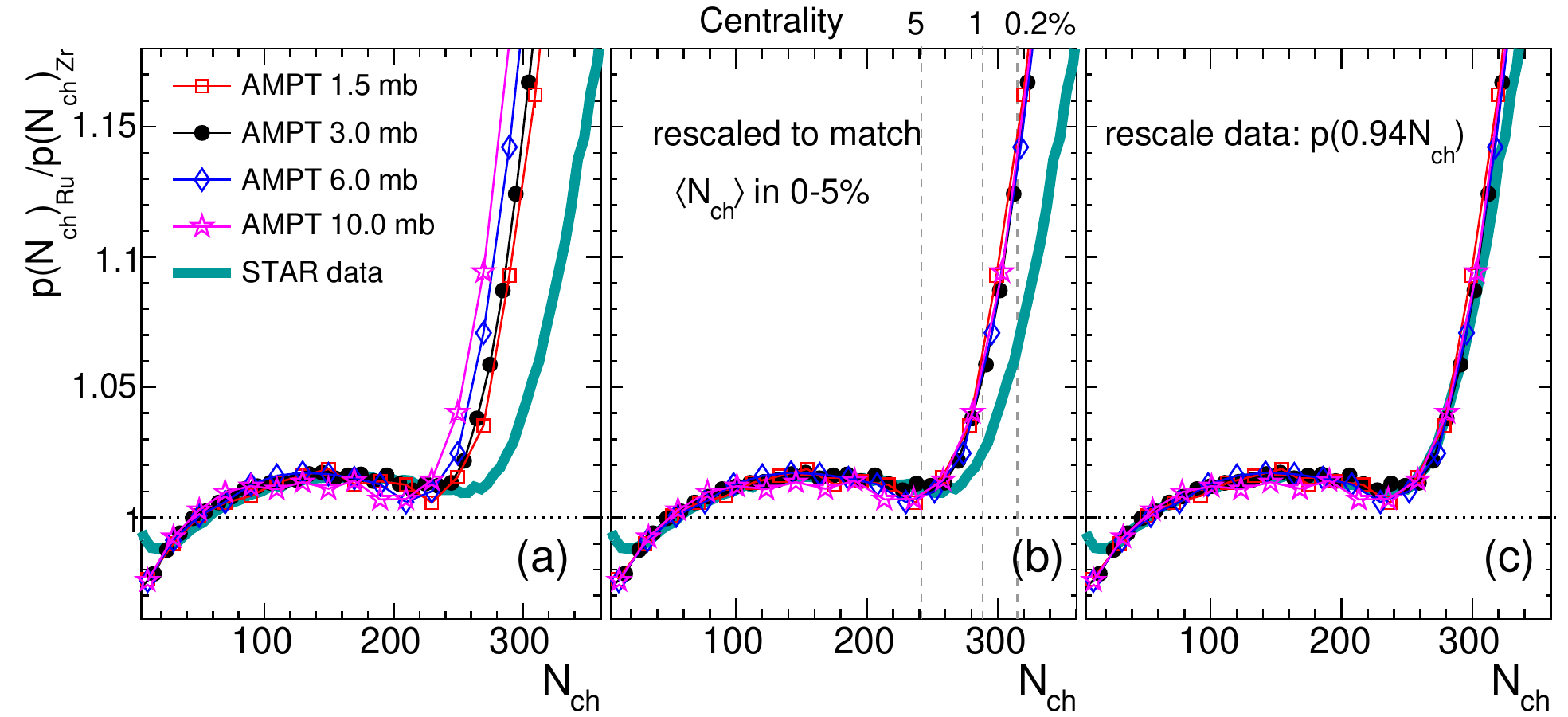}
\caption{\label{fig:1} (a): The $\nch$ dependence of the ratio $p(\nch)_{\mathrm{Ru}}/p(\nch)_{\mathrm{Zr}}$ for AMPT with different partonic cross sections listed in Tab.~\ref{tab:2} compared with the STAR data \cite{STAR:2021mii}. (b): All $p(\nch)$ distributions are rescaled to match the $\lr{\nch}$ value in 0-5\% centrality range of the data. The AMPT curves line up, but differ from the data. (c): The $x$-axis from the STAR data is further scaled by 0.94, i.e. $p(0.94\nch)_{\mathrm{Ru}}/p(0.94\nch)_{\mathrm{Zr}}$ in order to overlap with the AMPT.}
\end{figure*}

\textit{\bf Results.} We first study the final state effects on the most basic observable $p(\nch)$. Since the effect of viscosity is to enhance the entropy production during hydrodynamic evolution, the distribution $p(\nch)$ is expected to be broader for smaller partonic cross-section or larger viscosity. Table~\ref{tab:2} summarizes the average values of $\nch$ in 0--5\% central isobar collisions, compared with the experimental values.  We found that the $\nch$ value decreases by about 10\% from $\sigma=1.5$ $m$b to 10 $m$b independent of centrality, implying that the impact of $\eta/s$ is mostly an overall rescaling of $p(\nch)$. The table also lists the scale factor along the $x$-axis required to match the $\lr{\nch}$ of the data. 

Figure~\ref{fig:1} shows the isobar ratio $R_{\nch}=p(\nch)_{\mathrm{Ru}}/p(\nch)_{\mathrm{Zr}}$ for different cases. The ratios depend on the partonic cross-section, and they are also different from the data. But after applying the scale factors listed in Table~\ref{tab:2}, all ratios from the AMPT model collapse on to a single curve,  confirming that the effects of shear viscosity only amount to an overall rescaling of the $\nch$, but not the shape of $p(\nch)$. Yet, all the ratios still overshoot the data in the central region, and an additional scale factor, applied on the $x$-axis of the data by 0.94, is required to bring $R_{\nch}$ to match the STAR ratio. This implies that the shape of $p(\nch)$ and ratio $R_{\nch}$ from AMPT can not be matched to experimental data by a common scaling factor in $\nch$; the two factors required are off by about 6\%~\footnote{The $\nch$ in the experimental data is not corrected by tracking efficiency which usually has the form $\epsilon=a+b\nch$. This leads to a $\nch$-dependent rescaling of $p(\nch)$ and $R_{\nch}$ along the $x$-axis by an identical factor, which would not remove the discrepancy seen in the middle panel.}. This mismatch points to a non-trivial relation between collision geometry and the particle production mechanism, but exact mechanism requires more studies. In the following discussion, the first scale factors are always applied to the AMPT results before comparing with each other. 
\begin{figure}[h!]
\centering
\includegraphics[width=1\linewidth]{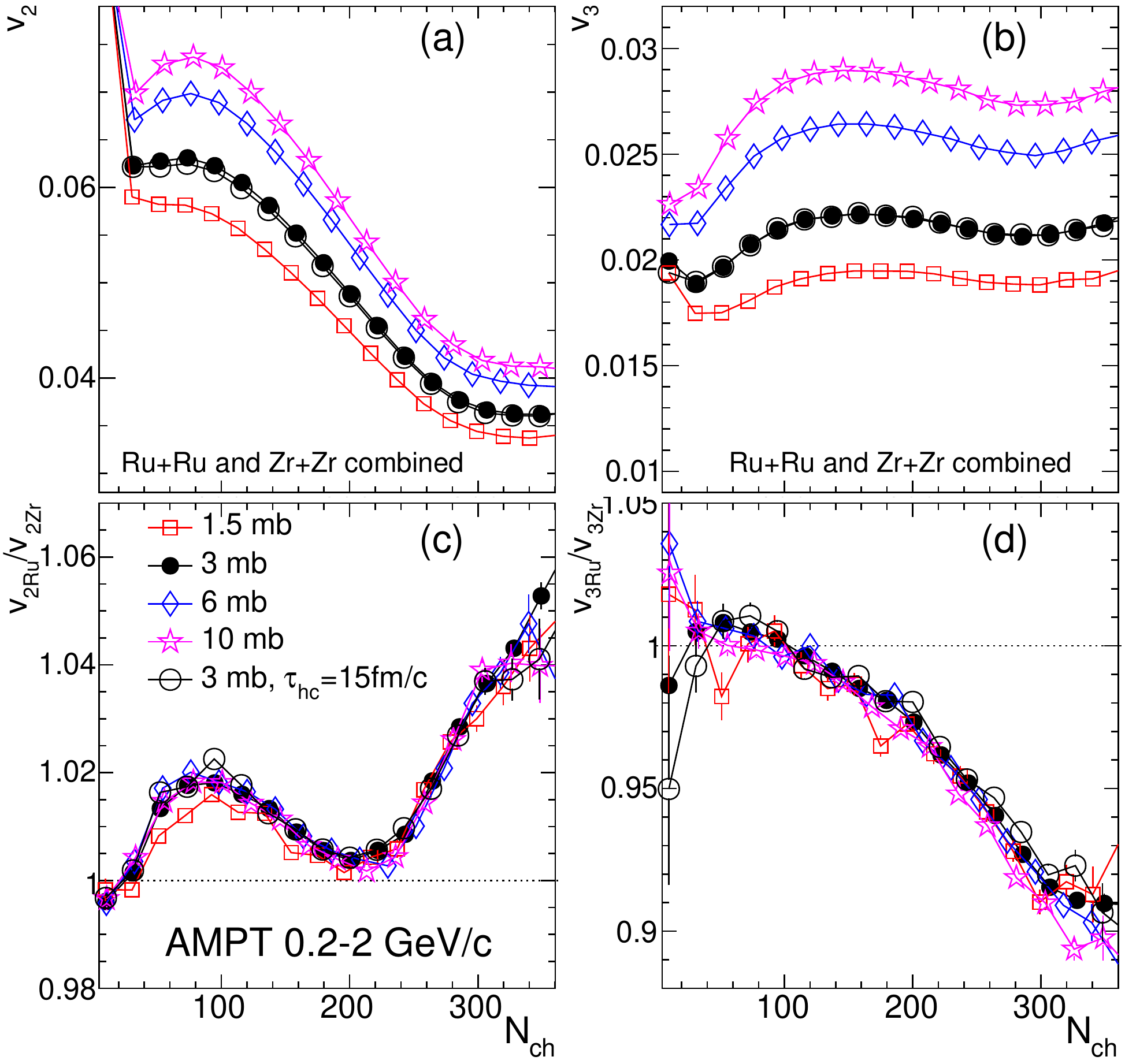}
\caption{\label{fig:2} The $\nch$ dependence of $v_2$ (left column) and $v_3$ (right column) by averaging the results from Ru+Ru and Zr+Zr collisions, labelled as ``Ru+Ru and Zr+Zr combined''  (top row) and the isobar ratios (bottom row) for different values of partonic cross-sections, using hadrons in $|\eta|<2$ and $0.2<\pT<2$ GeV/$c$. In the case of 3 $m$b, the result for changing hadronic cascade time to 15 fm/$c$ is also shown.}
\end{figure}

\begin{table}[h!]
\caption{\label{tab:2} The average of charged particle multiplicity, $\lr{\nch}_{0-5\%}$, in 0-5\% centrality from AMPT with different partonic cross sections and the data. The scale factors required to make the $\lr{\nch}_{0-5\%}$ approximately to match that of the data are given in the third row; and these factors are applied in both \ruru\ and \zrzr\ collisions. The last row gives the additional scale factor required to explicitly match the shape of $p(\nch)_{\mathrm{Ru}}/p(\nch)_{\mathrm{Zr}}$ between the data and AMPT.}
\begin{ruledtabular}
   \begin{tabular}{cccccc}
     & 1.5 mb & 3 mb& 6 mb& 10 mb & data\\
     \hline
Ru+Ru $\lr{\nch}_{0-5\%}$     & 281.60 & 274.36 & 265.17 & 258.43& 289.32\\
Zr+Zr $\lr{\nch}_{0-5\%}$     & 278.34 & 271.30 & 262.12 & 255.40& 287.36\\
scale factor (SF)            & 1.032  & 1.059  & 1.096  & 1.125 & 1\\
additional SF         & 1    & 1     & 1     &1     & 0.94\\
\end{tabular}
\end{ruledtabular}
\end{table}

In the picture of hydrodynamic, harmonic flow $v_2$ and $v_3$ are driven by the initial state eccentricities $\varepsilon_2$ and $\varepsilon_3$, respectively: $v_n = k_n \varepsilon_n$~\cite{Gardim:2011xv}. The response coefficients $k_n$, encoding all final state effects, are function of centrality but otherwise are the same for events in the same centrality class. The $k_n$ are also strong function of $\pT$ and depends on the mass of the hadrons. Therefore the sensitivity of the isobar ratio to the final state is reflected by the ratio of $k_n$ between \ruru\ and \zrzr\ collisions:
\begin{align}\label{eq:6}
\frac{v_{n,\mathrm{Ru}}}{v_{n,\mathrm{Zr}}} = \frac{k_{n,\mathrm{Ru}}}{k_{n,\mathrm{Ru}}}\frac{\varepsilon_{n,\mathrm{Ru}}}{\varepsilon_{n,\mathrm{Ru}}}\Longrightarrow R_{v_n} = R_{k_n} R_{\varepsilon_n}
\end{align}
One expect $R_{k_n} = k_{n,\mathrm{Ru}}/k_{n,\mathrm{Zr}}\approx 1$ for isobar collisions with the same $\nch$.

The impact of partonic cross-section and hadronic cascade time on $v_2$ and $v_3$ are studied in Fig.~\ref{fig:2}. Varying the cross-section changes the $v_2$ by up to 30\% and $v_3$ by up to 50\%. However the $R_{v_n}$ are almost unchanged, implying that the isobar ratios of harmonic flow are insensitive to the medium properties in the final state.  Only a small visible deviation of the scaling is observed in the mid-central collisions for $R_{v_2}$ in the 1.5 $m$b case. That is the region, where the $R_{v_2}$ is dominated by the nuclear structure parameter $a_0$~\cite{Jia:2021oyt}. Changing the hadronic cascade time has negligible impact on $v_n$ and $R_{v_n}$.
\begin{figure}[h!]
\centering
\includegraphics[width=1\linewidth]{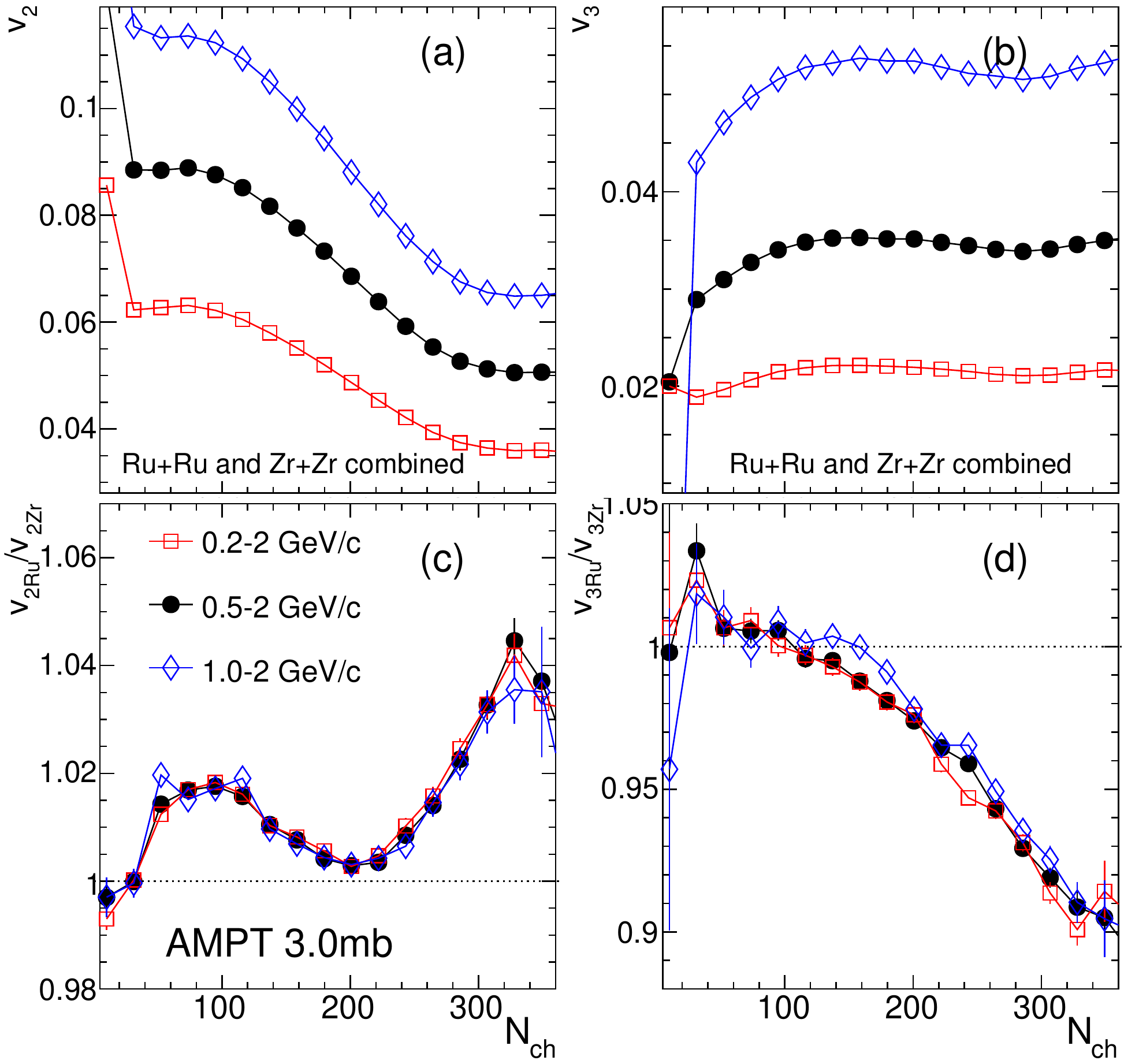}
\caption{\label{fig:3} The $\nch$ dependence of $v_2$ (left column) and $v_3$ (right column) in combined Ru+Ru and Zr+Zr collisions (top row) and the isobar ratios (bottom row) using hadrons in three $\pT$ ranges.}
\end{figure}

\begin{figure}[h!]
\centering
\includegraphics[width=1\linewidth]{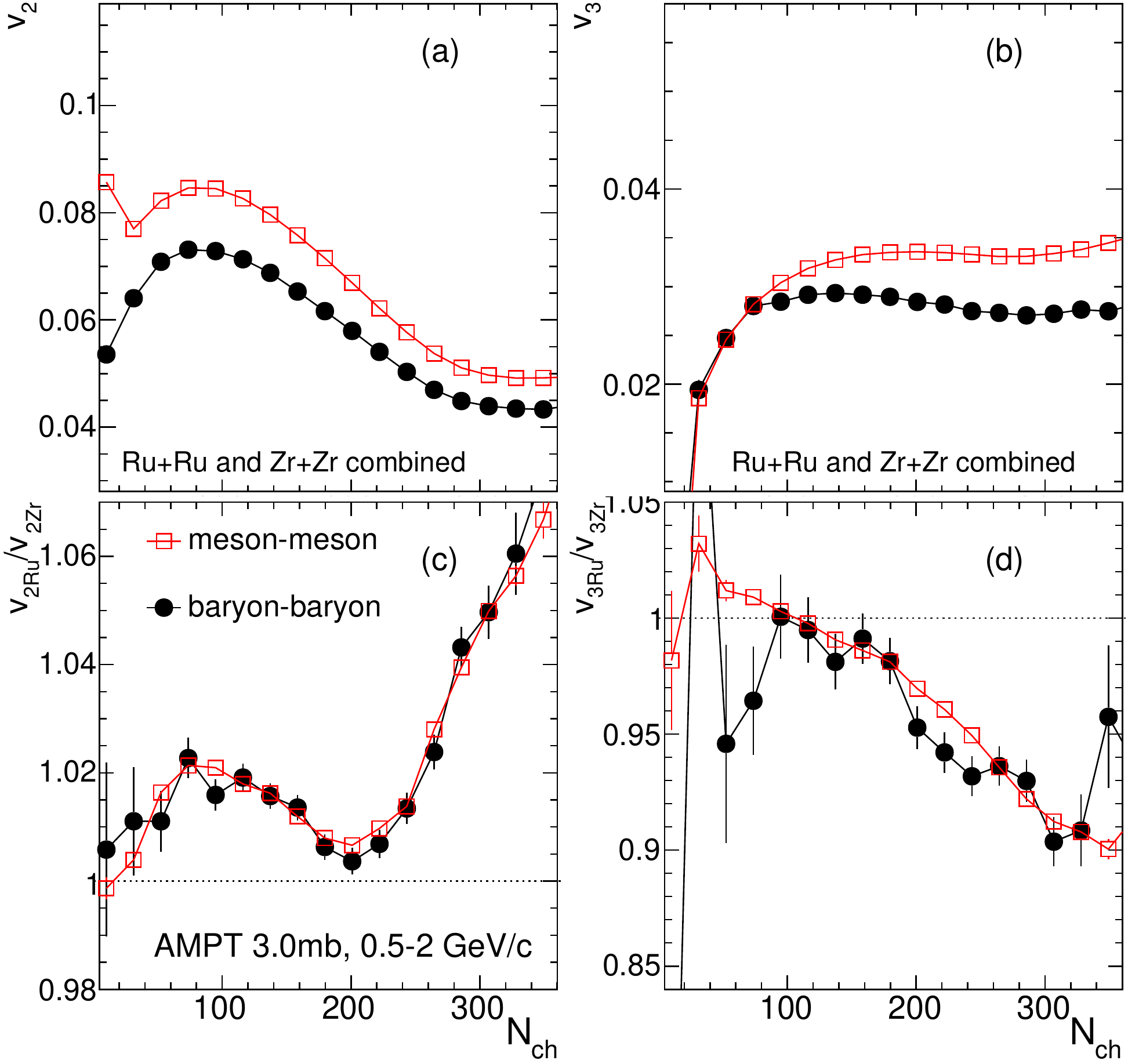}
\caption{\label{fig:4} The $\nch$ dependence of $v_2$ (left column) and $v_3$ (right column) in combined Ru+Ru and Zr+Zr collisions (top row) and the isobar ratios (bottom row) obtained using two-subevent method for mesons and baryons in $0.5<\pT<2$ GeV/$c$.}
\end{figure}

Figures~\ref{fig:3} and \ref{fig:4} display the $v_n$ and $R_{v_n}$ for three $\pT$ ranges and two hadron types, respectively. The values of $v_n$ are dramatically different among these different cases but the ratios are remarkably stable. The lack of $\pT$ dependence also implies that the $R_{v_n}$ are insensitive to the non-equilibrium effects such as the so-called $\delta f$ correction~\cite{Dusling:2009df}, while the similarity between meson and baryon implies that $R_{v_n}$ are insensitive to partonic coalescence at hadronization and hadronic transport~\footnote{Result in Fig.~\ref{fig:4} is obtained using two-suevent method where the particles in the pair are taken from different $\eta$ region. However, the $v_n$ ratios without $\Delta \eta$ gap do show some differences between meson-meson correlation and baryon-baryon correlation, suggesting that short-range correlations induced by parton coalescence may influence the PID dependence of $R_{v_n}$.}. 

\begin{figure}[h!]
\centering
\includegraphics[width=0.7\linewidth]{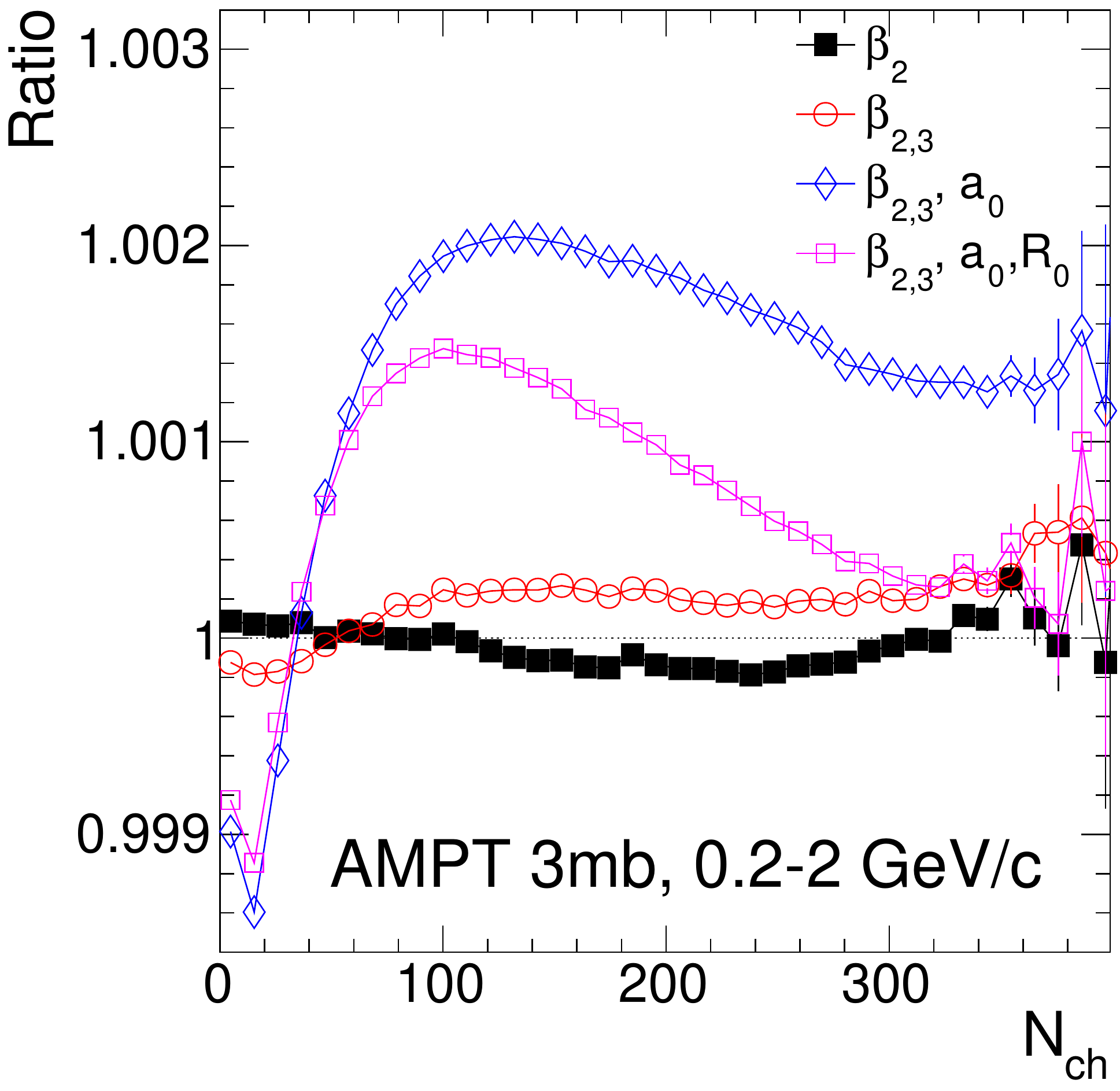}
\caption{\label{fig:5} The isobar ratio of $\lr{\pT}$ as a function of $\nch$ by incorporating the differences of four nuclear structure parameters between \ruru\ and \zrzr\ collisions one by one.}
\end{figure}

We also investigated the impact of nuclear structure on $\lr{\pT}$. We first performed a separate study where we isolate the contribution of each nuclear structure parameter similar to our previous study on $v_n$~\cite{Zhang:2021kxj}. As shown in Fig~\ref{fig:5}, $\beta_2$ and $\beta_3$ have little impact on the $R_{\lr{\pT}}$. In contrast, a significant enhancement of the $R_{\lr{\pT}}$ of about 0.2\% is observed in mid-central collision associated with the fact that $a_{0,\mathrm{Ru}}>a_{0,\mathrm{Zr}}$. But a large part of this enhancement, especially in central collisions, is cancelled by the fact that $R_{0,\mathrm{Ru}}<R_{0,\mathrm{Zr}}$. The enhancement of $R_{\lr{\pT}}$ after considering all four structure parameters produce a maximum enhancement of 0.15\% around $\nch=80$ or 35\% centrality. The centrality dependence trends are similar to the recent STAR preliminary measurement~\cite{chunjianhaojie}, but the magnitude is about a factor of three smaller. This is mainly due to the much weaker radial flow response to the size fluctuations in the AMPT model~\cite{Jia:2021qyu}. Thus the AMPT results shown here can only be used for qualitative guidance~\footnote{By appropriate choice of Glauber parameters, in particular with a large nucleon width, a previous study based on 2+1D VISNU hydrodynamic code were able to reproduce the experimental data quantitatively~\cite{Xu:2021uar}.}.

\begin{figure}[h!]
\centering
\includegraphics[width=1\linewidth]{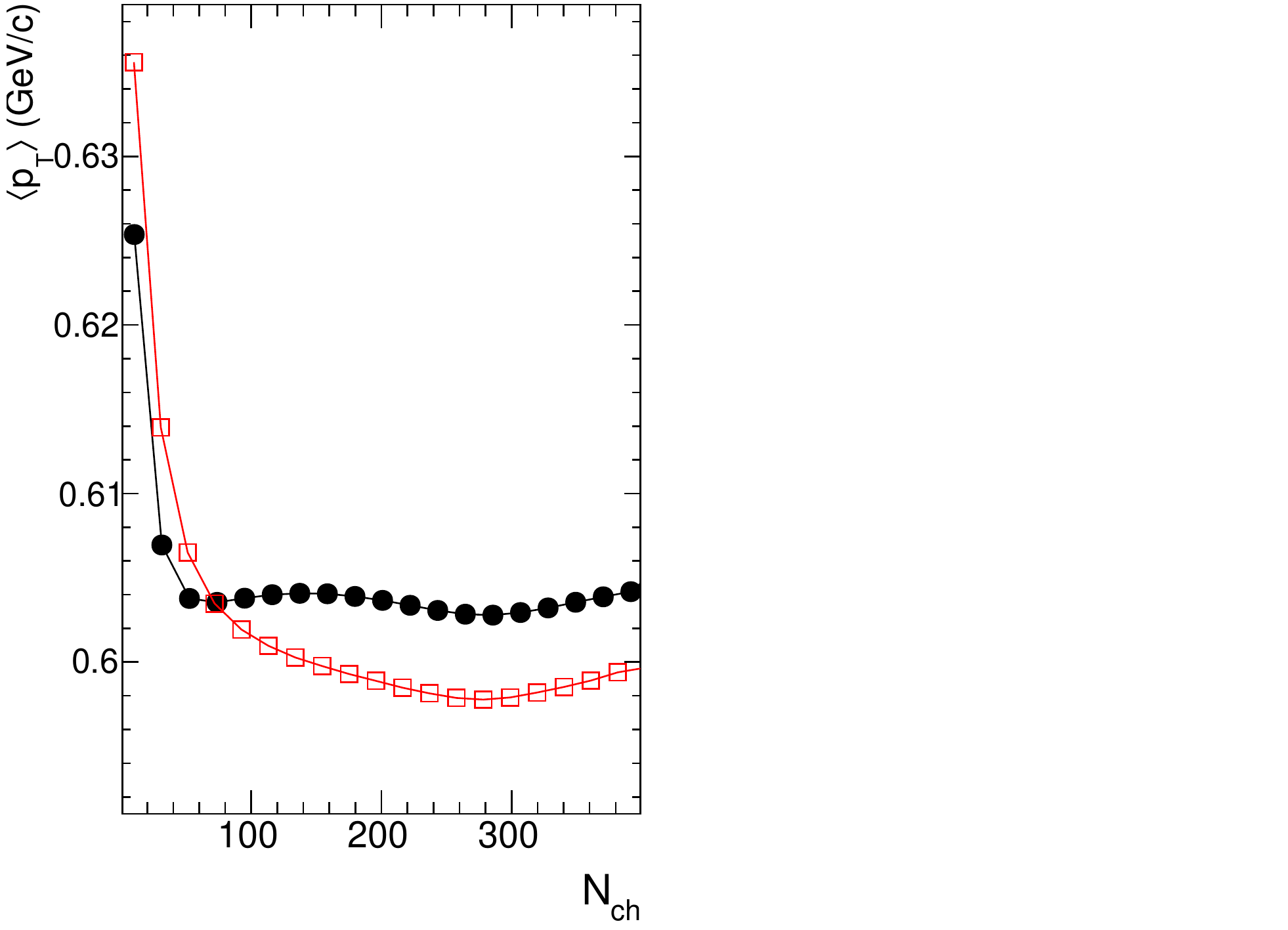}
\caption{\label{fig:6}  The $\nch$ dependence of $\lr{\pT}$ in combined Ru+Ru and Zr+Zr collisions (left) and the isobar ratio (right) for different values of partonic cross-sections. In the case of 3 $m$b, the result for hadronic cascade time of 15 fm/$c$ is also shown.}
\end{figure}

Figure~\ref{fig:6} displays the $\lr{\pT}$ and isobar ratio for different partonic cross-sections. Increasing the partonic cross-section increases the value of $\lr{\pT}$, and suggesting a stronger radial flow associated with larger pressure gradients. The $R_{\lr{\pT}}$ values are somewhat sensitive to viscosity, especially in the peripheral region. The apparent weaker dependence observed in a previous study~\cite{Xu:2021uar} is because that study only varies $\eta/s$ within 0.04--0.16, while our AMPT study varies $\eta/s$ in a much larger range of $0.09$--$0.39$. Clearly the nuclear structure influence on $\lr{\pT}$ is not entirely free of final state effects. However, the $\lr{\pT}$ is dominated by physics unrelated to QGP. In fact, only a small fraction, about 20\% of $\lr{\pT}$ in A+A collisions accounting for the increase of $\lr{\pT}$ from peripheral to central collisions, can be associated with genuine radial flow. This is different from $v_n$, which vanishes without final state interactions. So the modest sensitivity of $R_{\lr{\pT}}$ to viscosity may be attributed to modification of the component unrelated to collective flow by the final state interactions. On the other hand, changing the hadronic transport time from 30 fm/$c$ to 15 fm/$c$ influences the value of $\lr{\pT}$ but give nearly identical $R_{\lr{\pT}}$.

\textit{\bf Summary.} We studied the impact of final state interactions on the elliptic flow $v_2$, triangular flow $v_3$, mean transverse momentum $\lr{\pT}$ and charged particle multiplicity distribution $p(\nch)$ and the ratios between \ruru\ and \zrzr\ collisions at $\sqrt {s_{NN}}$ = 200 GeV within the AMPT framework. The deviation of these ratios from unity are related to difference of collective nuclear structure between Ru and Zr nucleus, in terms of quadrupole deformation $\beta_2$, octupole deformation $\beta_3$, nuclear skin $a_0$ and nuclear radius $R_0$.

We found that the distribution $p(\nch)$ is rescaled by a constant factor when varying shear viscosity, but the shape of ratio $p(\nch)_{\mathrm{Ru}}/p(\nch)_{\mathrm{Zr}}$ remains unchanged. The values of $v_{n}$ vary by up to 50\% when changing shear viscosity, but the ratios $v_{n,\mathrm{Ru}}/v_{n,\mathrm{Zr}}$ remain almost the same. Furthermore, the numerical values of $v_{n,\mathrm{Ru}}/v_{n,\mathrm{Zr}}$ are also found to be nearly identical between different $\pT$ ranges, as well as between the mesons and baryons. We also carried out a detailed investigation of the nuclear structure influence on the ratio of average transverse momentum $\lr{\pT}_{\mathrm{Ru}}/\lr{\pT}_{\mathrm{Zr}}$. This ratio is enhanced by the sharper skin of $^{96}$Ru i.e. $a_{0,\mathrm{Ru}}<a_{0,\mathrm{Zr}}$, but then is partially compensated by a larger radius of $^{96}$Ru i.e. $R_{0,\mathrm{Ru}}>R_{0,\mathrm{Zr}}$. On the other hand, the nuclear deformations have very little influence except in the ultra-central collision region. The values of $\lr{\pT}_{\mathrm{Ru}}/\lr{\pT}_{\mathrm{Zr}}$ are only slightly influenced by the viscosity and do not depend on the hadronic cascade time. These findings prove that the ratios of bulk observables in isobar collisions, in particular $v_{2}$ and $v_3$, are unique probe of the nuclear structure difference between isobar nuclei, and therefore provide a unique lever-arm to study the relation between nuclear structure and initial condition of the heavy ion collisions.

\textit{Acknowledgement.} We thank useful discussion with Giuliano Giacolone, Lumeng Liu, Zi-wei Lin, Wilke Van der Schee and Jun Xu. This work is supported by the U.S. Department of Energy under Grant No. DEFG0287ER40331.
\bibliography{finalstate}{}
\bibliographystyle{apsrev4-1}
\end{document}